\documentclass{article}

\usepackage{PRIMEarxiv}

\usepackage[utf8]{inputenc} 
\usepackage[T1]{fontenc}    
\usepackage{hyperref}       
\usepackage{url}            
\usepackage{booktabs}       
\usepackage{amsfonts}       
\usepackage{nicefrac}       
\usepackage{microtype}      
\usepackage{lipsum}
\usepackage{fancyhdr}       
\usepackage{graphicx}       
\graphicspath{{media/}}     
\usepackage{float}
\usepackage{tabularx}
\usepackage{longtable}
\usepackage{amsmath}
\usepackage{amssymb}
\usepackage[utf8]{inputenc}
\usepackage{textgreek}

\title{Inferring Factor Importance with Explainable AI Model for Macrophage Immune Cell Polarization
}

\author{
  Joseph Pateras \\
  Department of Computer Science \\
  Virginia Commonwealth University \\
  Richmond, VA, USA\\
  \texttt{\{paterasj@vcu.edu\}} \\
   \And
  Gurseerat Kaur\\
  Department of Computer Science \\
  Virginia Commonwealth University \\
  Richmond, VA, USA\\
  \And
  D. Joshua Cohen \\
  Department of Biomedical Engineering \\
  Virginia Commonwealth University \\
  Richmond, VA, USA\\
  \And
  Barbara D. Boyan\\
  Department of Biomedical Engineering \\
  Virginia Commonwealth University \\
  Richmond, VA, USA\\
  \And
  Zvi Schwartz \\
  Department of Biomedical Engineering \\
  Virginia Commonwealth University \\
  Richmond, VA, USA\\
  \And
  Preetam Ghosh\\
  Department of Computer Science \\
  Virginia Commonwealth University \\
  Richmond, VA, USA\\
}

\begin{document}
\maketitle

\begin{abstract}
Macrophages are highly plastic immune cells that dynamically shift between pro- and anti-inflammatory phenotypes in response to cytokines, transcriptional regulators, metabolic cues, and local microenvironmental conditions. These transitions shape immune homeostasis, chronic inflammation, cancer progression, biomaterial integration, foreign body responses, and tissue regeneration. Yet macrophage polarization is governed by highly nonlinear regulatory networks that are difficult to study experimentally across many physiological and pathological states. Here, we integrate mechanistic simulation, surrogate neural network modeling, and explainable machine learning to identify key regulatory drivers of polarization. We generated simulation data with a COPASI-based macrophage polarization model and used them to train an LSTM surrogate capable of reconstructing temporal macrophage marker dynamics across diverse macrophage activation states. SHAP, Individual Conditional Expectation, and Partial Dependence Plot analyses quantified feature importance, identified nonlinear response regimes, and characterized higher-order interactions among cytokines, transcription factors, and polarization markers. The model identified oxygen availability and IL10 as dominant anti-inflammatory regulators, consistently promoting M2-associated outputs such as ARG1 and VEGF. Conversely, IFN-$\gamma$, TNF$\alpha$, IL12, iNOS, CXCL9, and mCXCL10 reinforced M1-associated inflammatory programs. STAT1-centered analyses identified iNOS, TNF$\alpha$, and mCXCL10 levels as strong contributors to STAT1 outcomes, consistent with amplification of the IFN-$\gamma$/JAK/STAT1 inflammatory axis. VEGF negatively contributed to predicted STAT1 output, suggesting antagonism between repair-associated M2 markers and M1 transcriptional activation. SHAP, ICE, and PDP analyses also showed that STAT1 regulation is nonlinear and threshold-like: pro-inflammatory signals must accumulate before STAT1 output rises sharply. Together, the results demonstrate that the COPASI-informed LSTM explainability framework can identify polarization drivers, uncover nonlinear regulatory thresholds, and prioritize pathways for experimental validation in immune imbalance, chronic inflammation, cancer, and implant-associated foreign body response.
\end{abstract}


\section{Introduction}
\subsection{Computational Immune Response Modeling}
\subsubsection*{Macrophage Biology}
Macrophages are central regulators of immune decision-making across infection, bone regeneration, tissue injury, cancer, and biomaterial implantation. Rather than functioning as static cell types, macrophages continuously interpret biochemical, mechanical, metabolic, and cellular signals from their local microenvironment to coordinate inflammation, repair, angiogenesis, matrix remodeling, and immune suppression. This context-dependent plasticity makes macrophages both powerful mediators of immune defense. As a result, macrophage phenotype switching has emerged as a critical target for precision medicine, where patient- or disease-specific models may help predict immune trajectories, identify dominant regulatory mechanisms, and guide therapeutic interventions that restore appropriate immune balance.

Macrophages capable of adopting distinct polarization states, primarily named M1 (pro-inflammatory) and M2 (anti-inflammatory). This dynamic behavior is simplified into two polarized states: M1 macrophages drive inflammatory responses critical for pathogen elimination and tissue defense, while M2 macrophages support tissue repair, resolution of inflammation, and immune regulation \cite{murray2014m1m2, mantovani2004polarization, mantovani2009polarization}.

\subsubsection*{Previous Computational Models} A deeper understanding of the mechanisms governing macrophage polarization is therefore essential for developing targeted therapies and biomedical interventions for diseases characterized by immune dysfunction, such as chronic inflammatory conditions, autoimmune disorders, cancer, and implant rejection. Computational modeling offers a powerful approach for unraveling the complex signaling networks that regulate macrophage polarization. The comprehensive model developed by Zhao et al.~\cite{zhao2019model} integrates multiple pathways, including the IFN-$\gamma$, IL-4, and hypoxia signaling cascades, to simulate macrophage behavior under various conditions.

The complexity of macrophage polarization is increasingly appreciated through computational and systems biology approaches. For instance, Zhao et al.~\cite{zhao2019model} demonstrated that modeling of immune networks can reveal emergent behavior in cytokine regulation and macrophage phenotype switching. Model sensitivity analyses can reveal key molecular nodes and interactions as apt targets for potential therapeutic interventions for the pathophysiology of arterial disease and cancers. Further emphasizing the tumor microenvironment context, Leon et al.~\cite{leon2023uncoding} used network-based methods to decode the interdependency between macrophage behavior and external environmental signals, reinforcing the idea that external cues profoundly modulate polarization states. Studies such as Zhang et al.~\cite{zhang2022angiogen} underscore the utility of dynamic simulations and machine learning models in mapping immune phenotypes and identifying potential therapeutic leverage points across various cancers. Wang et al.~\cite{wang2022brca} developed a quantitative systems pharmacology model that incorporates tumor-associated macrophage (TAM) dynamics to predict immunotherapy outcomes in triple-negative breast cancer. By integrating TAM polarization, cytokine profiles, and digital pathology data, their model captures the immunosuppressive impact of M2 macrophages and supports macrophage-targeted therapeutic strategies. Also studying TAMs, Cess et al. \cite{cess2020TAM} model the TAM/Tumor Cell/T-Cell crosstalk to determine how the immune response changes due to three macrophage-based immunotherapeutic strategies: macrophage depletion, recruitment inhibition, and macrophage reeducation; concluding that reeducation is the best strategy, while TAM recruitment and tumor proliferation play an important role. Torres et al. \cite{torres2019immune} developed and fit a mathematical model of immune-cell influx during mouse peritonitis, including simplified macrophage polarization, then evaluated model robustness using identifiability and sensitivity analyses. The model reproduced expected inflammatory timing and identified apoptotic neutrophils as a major driver of the inflammatory response. Simulated interventions suggest the model can be used to test macrophage-targeted therapies before experimental validation. Guo et al. \cite{guo2025sc} represents an important convergence of single-cell imaging, and mathematical modeling toward precision medicine. By combining live-cell measurements of NFκB dynamics with simulations that extend beyond experimentally accessible conditions, the work shows how heterogeneous macrophage responses can be integrate at both the population and single-cell levels. Zerrouk et al. \cite{zerrouk2024ra} offer a study which highlights the use of mechanistic models toward developing new therapeutic targets. For modeling the M1/M2 macrophage ratio in the context of rheumatoid arthritis, this study employs knockout experiments to identify NFkB, JAK1/JAK2 and ERK1/Notch1 as targets for suppressing pro-inflammatory macrophages, and GSK3B as a promoter for anti-inflammatory macrophages. For responding to foreign body penetration, Anderson et al. \cite{anderson2008foreign} conclude that ersistent M1 polarization can drive chronic inflammation an granulation or fibrotic encapsulation, while M2 polarization is associated with tissue remodeling, wound healing, and improved integration of biomaterials \cite{anderson2008foreign}.

Taken together, this collection of studies highlight macrophage polarization as a broadly applicable modeling target across inflammatory disease, cancer, and immune-mediated tissue remodeling. The convergence of mechanistic mathematical modeling, systems pharmacology, machine learning, single-cell imaging, and dynamic signaling analysis provides an increasingly rich framework for understanding macrophage behavior across biological scales.

Integrating dynamic macrophage models for the development of disease-specific and patient-informed pharmacological models offers a promising strategy for identifying therapeutic leverage points, testing interventions in silico, and ultimately designing precision immunomodulatory strategies tailored to specific environments.

\subsection{Explainability}
\subsubsection*{Surrogate Modeling}
The polarization process is regulated by a large network of interacting genes, cytokines, transcription factors, and microenvironmental signals. Because these components jointly determine whether macrophages adopt pro-inflammatory M1-like, anti-inflammatory M2-like, or intermediate phenotypes, it is essential to quantify their relative importance in maintaining polarization balance and driving phenotype transitions.

To address this challenge, a relative gene-importance modeling study was conducted using a mechanistic computational model of macrophage polarization. Simulation data were generated in COPASI (COmplex PAthway SImulator) \cite{hoops2006copasi} and used to train surrogate machine learning models. COPASI is a widely adopted platform for simulating biochemical reaction networks, enabling time-dependent molecular behavior to be modeled through reaction rate equations and concentration-based dynamics. 

This surrogate modeling strategy is particularly useful because experimental macrophage studies are constrained by the number of conditions that can be evaluated in vivo or in vitro. In biological experiments, systematically varying many cytokines, transcription factors, pathway activities, and microenvironmental cues across multiple doses and time points is often infeasible due to cost, sample availability, biological variability, and ethical limitations. By contrast, computational simulation allows the model to explore an experimentally impractical number of pathological and therapeutic states, including combinations of perturbations that would be difficult or impossible to generate directly in the laboratory.

In this study, COPASI was used to simulate macrophage polarization across diverse biological conditions relevant to the IFN-$\gamma$, IL-4, and hypoxia signaling pathways. These simulations included variation in cytokine levels, transcription factor activity, and microenvironmental stimuli, producing time-series data for molecular species involved in M1 and M2 polarization. The generated datasets captured balanced polarization states as well as perturbed, skewed, and disease-relevant phenotypes. Surrogate machine learning models trained on these simulations can then rapidly approximate the behavior of the underlying macrophage model, enabling systematic gene-level perturbation studies and feature-importance analyses across a much larger state space than would be experimentally accessible. In this way, surrogate modeling complements experimentation by using mechanistic simulations to identify candidate regulatory drivers, prioritize genes or pathways for validation, and generate hypotheses about macrophage behavior under complex pathological conditions. Full model details, summarized in Table~\ref{tab:model}, are provided by Zhao et al.~\cite{zhao2019model}.

\subsubsection*{SHAP Feature Attribution}
Advancements in interpretable machine learning, such as laid out in Lundberg et al.~\cite{lundberg2020shap}, provide tools like the SHapley Additive exPlanations (SHAP) to dissect feature contributions in black-box models, which are increasingly applied to biological network analysis. The integration of these methods with simulation frameworks allows for mechanistic interpretation of gene contributions, which can be validated against known markers and experimental perturbations.

\subsubsection*{ICE and Partial Dependence Analysis}
Complementary to SHAP, feature effect methods such as Individual Conditional Expectation (ICE) and Partial Dependence Plots (PDP) \cite{goldstein2015icepdp} provide insight into how model predictions respond to variations in specific inputs. ICE curves capture sample-specific prediction trajectories by varying a single feature while holding others constant, thereby revealing heterogeneity and interaction effects that are often obscured in global summaries. PDPs summarize the average influence of a feature across the data distribution. Together, these approaches enable identification of nonlinear behaviors, threshold effects, and regime-dependent dynamics in gene regulatory networks. When applied to temporal or mechanistic models, ICE and PDP analyses further allow interpretation of how perturbations propagate through the system, offering a functional perspective that complements the attribution-based insights provided by SHAP.

Through the integration of computational simulations, explainable machine learning, and data-driven modeling, this study offers insights into the relative importance of regulatory mechanisms for macrophage polarization and their therapeutic implications. 

\subsubsection*{Biomaterial Integration Model Specificity} Following implantation, macrophages are among the first immune cells to interact with the foreign material surface, where their phenotype influences whether the local environment progresses toward constructive tissue integration, chronic inflammation, fibrotic encapsulation, or biomaterial failure. Persistent M1 polarization can drive chronic inflammation, granulation, and fibrotic encapsulation, while M2 polarization is associated with tissue remodeling, wound healing, regeneration and improved integration of biomaterials \cite{anderson2008foreign}. Macrophage responses are influenced by implant material type, and physical properties such as size, pore structure, and surface texture. Sterilization methods can also affect these responses by affecting lasting biochemical changes or leaving chemical residues \cite{boersema2016implant}. Secretions of the macrophage cell are sensitive to features like material topography \cite{chen2010topograph}.Understanding and controlling macrophage polarization has become one of the key objectives in the design of next-generation biomaterials, implants, and tissue engineering scaffolds. In titanium implants, this response is further complicated by interactions with mesenchymal stem cells, bone-forming osteoblasts, bone-resorbing osteoclasts, gingival fibroblasts, periodontal ligament cells, endothelial cells, and other resident or recruited immune cells, as well as by local inflammatory and microbial stimuli. Therefore, some factors may primarily reflect classical inflammatory activation, while others may be more directly related to tissue remodeling, osteointegration, angiogenesis, foreign body response, or communication with bone-resident cells.

For this reason, the scanned parameters include both canonical macrophage polarization regulators and biomaterial-relevant environmental features. Pro-inflammatory mediators such as IFN-$\gamma$, TNF$\alpha$, IL12, iNOS, CXCL9, and mCXCL10 are included because they may indicate sustained M1-like activation, which can promote immune cell recruitment and inflammatory signaling around the biomaterials. In contrast, IL4, IL10, ARG1, VEGF, STAT6, IRF4, and PPAR$\gamma$ represent regulatory, repair-associated, or M2-like pathways that may support vascularization, matrix remodeling, and resolution of inflammation. These features are particularly relevant in titanium implant systems in bone, where successful osseointegration requires a balanced immune response that permits bone formation rather than persistent inflammatory degradation or fibrosis. By scanning these parameters systematically, the surrogate model can identify which macrophage genes are most predictive of favorable or unfavorable immune states. This provides a mechanistic basis for determining which biological factors should be prioritized in future experimental studies.

\section{Methods}

\subsection{Simulation Dataset Generation}
To explore macrophage regulatory mechanisms relevant to biomaterial environments, we constructed simulation datasets that systematically toggle key signaling inputs associated with the foreign body response. 

Using the COPASI macrophage polarization model from Zhao et al.~\cite{zhao2019model}, we generated simulation conditions in which pathway inputs were selectively varied. Specifically, datasets were constructed by modulating signaling drivers associated with inflammatory activation, including IFN-$\gamma$/JAK--STAT1, and IL12/STAT4 pathways; tissue-repair and wound-healing responses, including IL4/IL4R$\alpha$--STAT6--IRF4, IL10/STAT3, ARG1-associated arginine metabolism, and VEGF/VEGFR2 signaling; and environmental stress responses such as hypoxia/HIF-1$\alpha$ signaling.These toggled inputs allow the simulations to capture macrophage behavior across a spectrum of immune responses, ranging from strongly inflammatory to pro-regenerative states.

By systematically varying these regulatory inputs across simulations, we produced datasets that reflect diverse immune microenvironments encountered during implant integration and foreign body response. This design enables downstream machine learning models to identify genes and regulatory pathways that most strongly influence macrophage polarization outcomes under implant-relevant conditions.

The training set will consist of $50,000+$ unique combinations of environmental, regulatory, or transcription factors and the corresponding time course of the polarization system. This diverse dataset will enable the downstream analysis in two major ways: SHAP values allow us to rank the relative importance between the outcome of one feature and the conditions prescribed by each of the others, and extrapolate the model to unseen states and explore pathologically preferred states.

\subsection{Mechanistic Model Summary} 
\begin{table}[h!]
\centering
\resizebox{\textwidth}{!}{%
\begin{tabular}{|l|c|c|c|}
\hline
\textbf{Model Component} & \textbf{Quantity} & \textbf{Description} & \textbf{Parameterization} \\
\hline
Reactions (fluxes) & 140 & IL-4, IFN-$\gamma$, hypoxia, feedback loops & Fitted or estimated from literature  \\
\hline
Unique species & 34 & Core molecules (cytokines, TFs, enzymes, mRNAs) & 31 species calibrated to literature data \\
\hline
Model nodes & 109 & Proteins, RNAs, complexes, derived from unique species & Initial copy numbers in molecules per cell (1 pL) \\
\hline
Parameters & 141 & Rate constants, affinities, degradation rates & Combination of fitted + published data \\
\hline
Calibration datasets & $>$40 & Time-course or steady-state measurements & IFN-$\gamma$, IL-4, hypoxia modules \\
\hline
Pathways covered & 3 major modules & IFN-$\gamma$ signaling, IL-4 signaling, hypoxia response & Includes crosstalk (e.g., STAT1/STAT6, HIF) \\
\hline
\end{tabular}%
}
\caption{Summary of model complexity integrating reactions, species, and calibration data.}\label{tab:model}
\end{table}

\begin{table}[h!]
\centering
\resizebox{0.95\textwidth}{!}{%
\begin{tabular}{|l|l|}
\hline
\textbf{Category} & \textbf{Unique Species (basis set)} \\
\hline
Cytokines & IL4, IFN$\gamma$, IL10, IL12, TNF$\alpha$, VEGF \\ 
\hline
Receptors & IL4R, IFNGR \\
\hline
Kinase Signaling Core & JAK, AKT \\
\hline
STAT Transcription Factors & STAT1, STAT6 \\
\hline
SOCS Regulators & SOCS1, SOCS3 \\
\hline
Interferon Regulatory Factors & IRF1, IRF4, IRF9 \\
\hline
Hypoxia / Oxygen Sensing & HIF1$\alpha$, HIF2$\alpha$, HIF1$\beta$, PHD, O$_2$ \\
\hline
Downstream Metabolic Machinery & IRG1, Itaconate, PTEN \\
\hline
miRNA Layer (post-transcriptional control) & miR93, miR3473, mPTEN (mRNA), mIRF9 (mRNA) \\
\hline
Immune Effector Enzymes & iNOS, ARG1 \\
\hline
Nuclear Hormone Receptor & PPAR$\gamma$ \\
\hline
Chemokine Outputs & CXCL9, mCXCL10 (mRNA) \\
\hline
\end{tabular}%
}
\caption{The 34 functionally–unique species that form the basis from which all 80 ODE model nodes are derived.\label{tab:unique}}
\end{table}

From the 34 functionally unique basis species, the largest expansion in node count arises from the shared signaling hub JAK (18 derived nodes), which reflects its combinatorial engagement across both IL--4 and IFN--$\gamma$ receptor complexes. STAT1 and IL4R are the next most expansive (9 derived nodes each), owing to multiple phosphorylation, dimerization, trafficking, and nuclear shuttling states; IL4 (8 nodes), IFNGR and STAT6 (7 nodes each), and IFN--$\gamma$ (6 nodes) follow. Moderate expansion occurs for core feedback regulators (SOCS1 and SOCS3: 5 nodes each) and for hypoxia transcriptional factors (HIF1$\alpha$, HIF2$\alpha$, and IRF9: 4 nodes each). In contrast, classical effector molecules appear as single-state outputs (ARG1, iNOS, IL12, TNF$\alpha$, CXCL9, VEGF, mCXCL10). Consistent with macrophage polarization biology, expansion targets associated with pro-inflammatory M1 programming (e.g.\ IFN--$\gamma$, STAT1, IRF1, iNOS, IL12, CXCL9, mCXCL10) tend to expand through distinct signal-processing cascades \cite{mills2000m1}, while M2--associated regulators (e.g.\ IL4, STAT6, IRF4, ARG1, IL10, PPAR$\gamma$, VEGF) expand through IL--4--centric alternative-activation modules \cite{mantovani2004m2}.
\subsection{Surrogate Model}

\begin{figure}[h!]
    \centering    \includegraphics[width=\linewidth]{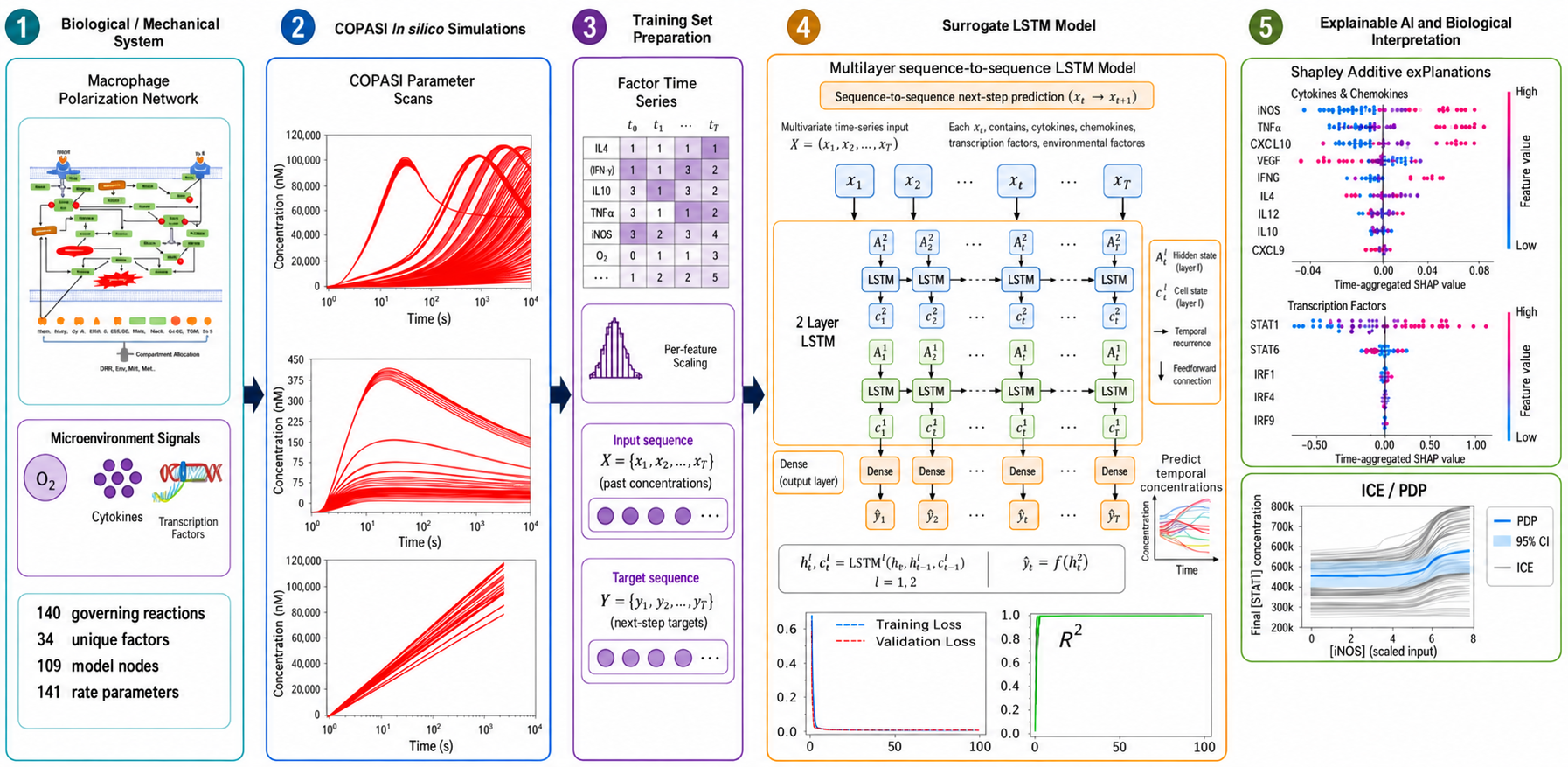}
    \caption{Diagram of the full surrogate LSTM workflow used to model the temporal dynamics of the macrophage system.}
    \label{fig:model}
\end{figure}

The workflow and model used to train this data is summarized in Figure \ref{fig:model}. We employed a Long Short-Term Memory (LSTM) architecture to learn the full temporal evolution of cytokine and transcription factor concentrations. The model operates on time-resolved input sequences and predicts the corresponding sequence of downstream states, enabling the capture of dynamic regulatory interactions and temporal dependencies inherent to macrophage polarization. Unlike static feedforward models, this framework learns the evolution of system state over time, enabling prediction of full concentration trajectories rather than terminal outputs alone.

\subsubsection*{LSTM Architecture}
Each sample is represented as a multivariate time series
\[
\mathbf{X} = \{\mathbf{x}_1, \mathbf{x}_2, \dots, \mathbf{x}_T\}, \quad \mathbf{x}_t \in \mathbb{R}^{d},
\]
where $\mathbf{x}_t$ contains cytokine, chemokine, transcription factor, and environmental features at time $t$. The model is trained in a sequence-to-sequence (next-step prediction) setting, learning the mapping
\[
\mathbf{x}_t \rightarrow \mathbf{x}_{t+1}.
\]

The temporal evolution of the system is encoded using an LSTM:
\[
\mathbf{h}_t, \mathbf{c}_t = \mathrm{LSTM}(\mathbf{x}_t, \mathbf{h}_{t-1}, \mathbf{c}_{t-1}),
\]
where $\mathbf{h}_t$ is the hidden state and $\mathbf{c}_t$ is the cell state capturing long-term dependencies. The LSTM acts as a latent dynamical encoder, integrating historical signaling information and capturing time-dependent regulatory effects.

At each time step, the latent state is mapped to predicted outputs via a feedforward neural network (the latent neural network, LNN):
\[
\hat{\mathbf{y}}_t = f_{\theta}(\mathbf{h}_t),
\]
where $f_{\theta}$ consists of fully connected layers with nonlinear activation functions. This mapping translates the learned latent representation into observable concentrations of target molecular species.

The model is trained to minimize the discrepancy between predicted and true trajectories across all time steps using an $L^1$ loss:
\[
\mathcal{L} = \frac{1}{T} \sum_{t=1}^{T} \left\| \hat{\mathbf{y}}_t - \mathbf{y}_t \right\|_1.
\]

This formulation enables the model to learn a latent dynamical system in which the hidden state $\mathbf{h}_t$ represents the evolving regulatory state of the macrophage system. By incorporating temporal context, the LSTM captures history-dependent effects, nonlinear interactions, and delayed responses that are not accessible to static models. This is particularly important for immune signaling, where prior exposure to cytokines and transcriptional regulators influences subsequent system behavior.

The learned latent representation provides a basis for downstream interpretability analyses, including SHAP-based feature attribution and ICE/PDP-based functional interrogation, allowing both global importance and context-dependent interactions to be quantified within a unified modeling framework.

\subsubsection{Workflow Summary}
The predictive model was implemented in PyTorch as a sequence-to-sequence architecture. Input data consisted of multivariate time series of cytokine and transcription factor concentrations, which were standardized using a zero-mean, unit-variance transformation across all timepoints. The model comprises a multi-layer Long Short-Term Memory (LSTM) network with hidden dimension 128, followed by fully connected layers that map latent temporal representations to predicted output trajectories.

Training was performed using the Adam optimizer (learning rate = $1\times10^{-4}$) with an $L^1$ loss function applied across all predicted timepoints. Gradient clipping was employed to stabilize training. Models were trained for 75 epochs with a batch size of 32, using a 70/30 train–test split. Model performance was evaluated using mean absolute error and $R^2$ metrics computed both across the full predicted sequence and at the final timepoint. Training and validation losses, along with per-target $R^2$ scores, were tracked throughout training to monitor convergence and predictive fidelity.

Feature importance and interaction effects were analyzed using SHAP values computed via \textbf{GradientExplainer}, alongside ICE and PDP analyses. Together, these methods provide complementary perspectives: SHAP quantifies global feature relevance, while ICE/PDP analyses reveal the structure of learned nonlinear relationships and higher-order interactions within the modeled biological system.

\section{Results}

\subsubsection*{Surrogate LSTM Model Performance}

The surrogate models are trained using data generated via COPASI parameter scans. Key parameters—including interferon-$\gamma$ concentration, IFN-$\gamma$ ([IFNG]$_0$), IL-12 ([IL12]$_0$), and other cytokine inputs—are co-varied across wide biologically relevant ranges to create a heterogeneously distributed training set spanning diverse macrophage activation states.

To assess model performance, an LSTM model was trained to predict all thirteen outputs. Figure \ref{fig} shows rapid convergence of both training and test loss. The corresponding $R^2$ scores for each predicted target also quickly approach 1.0, indicating that the model reliably reconstructs the simulated temporal dynamics and captures the underlying regulatory structure of the system.

\begin{figure}[h!]
\centering
\includegraphics[width=0.8\linewidth]{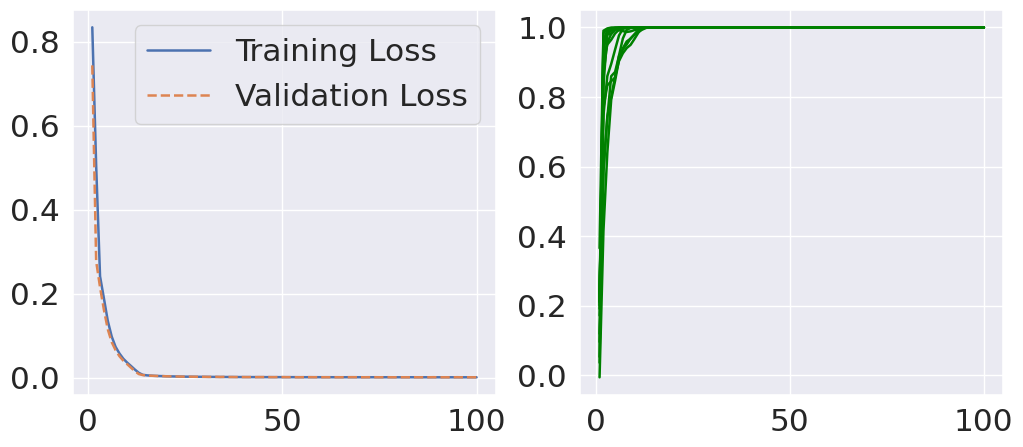}
\caption{Training and validation loss (left) and $R^2$ scores (right) for the neural network model predicting thirteen marker feature trajectories.}
\label{fig}
\end{figure}

\subsubsection*{SHAP Feature Attribution}

To investigate the contribution of individual input features to the model output, SHAP values were computed for each predicted molecular species. STAT1 is presented as a representative example because it is a central regulator of inflammatory macrophage activation. The corresponding SHAP summary plots are shown in Figure \ref{fig}.

\begin{figure}[h!]
\centering
\includegraphics[width=0.9\linewidth]{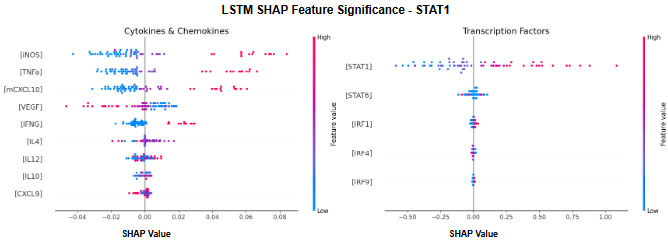}
\caption{SHAP summary plots for STAT1 prediction using the LNN–LSTM model. Left: cytokines and chemokines. Right: transcription factors. Features are ranked by mean absolute SHAP value, with color indicating relative feature magnitude. Pro-inflammatory mediators (iNOS, TNFa, mCXCL10) and STAT1 itself emerge as dominant drivers, highlighting both upstream signaling and autoregulatory effects.}
\label{fig}
\end{figure}

The SHAP analysis reveals that STAT1 prediction is strongly influenced by both inflammatory effector molecules and transcriptional regulators. Among the cytokine and chemokine features, iNOS, TNF$\alpha$, and mCXCL10 exhibit the largest absolute SHAP values. High values of these features generally contribute positively to predicted STAT1 concentrations, consistent with coordinated activation of the IFN-$\gamma$/JAK/STAT1 inflammatory program. In contrast, high VEGF values produce predominantly negative contributions, indicating an inverse relationship between reparative or pro-angiogenic signaling and predicted STAT1 activity.

Within the transcription factor group, STAT1 itself has the largest SHAP contribution, reflecting the strong autoregulatory dynamics represented in the mechanistic model and learned by the surrogate. STAT6 displays a broad bidirectional distribution of SHAP values, consistent with context-dependent competition between inflammatory and alternative activation programs. IRF1, IRF4, and IRF9 have smaller SHAP magnitudes, suggesting that their predictive roles are secondary or become prominent only under particular molecular conditions.

The spread and asymmetry of the SHAP distributions further indicate that these relationships are nonlinear. High levels of the dominant inflammatory regulators can produce disproportionately large increases in predicted STAT1 output, while the direction and magnitude of several lower-ranked features depend on the surrounding cytokine and transcription-factor environment.

\subsubsection*{ICE and Partial Dependence Analysis}

Although SHAP identifies which features most strongly contribute to STAT1 prediction, it does not by itself show how the predicted response changes across the range of each feature. Individual conditional expectation (ICE) curves and partial dependence plots (PDPs) were therefore used to examine the learned response functions for selected high-ranking regulators.

The ICE curves illustrate the response of predicted STAT1 concentration to perturbations in individual features, while the PDP represents the population-averaged effect with associated confidence intervals. These results are shown in Figure \ref{fig}.

\begin{figure}[h!]
\centering
\begin{tabular}{c c c}
\includegraphics[width=0.31\linewidth]{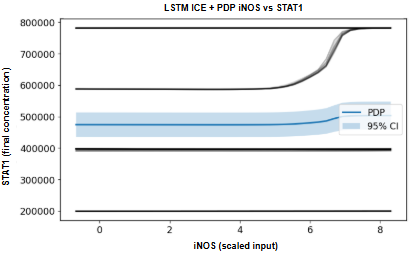} &
\includegraphics[width=0.31\linewidth]{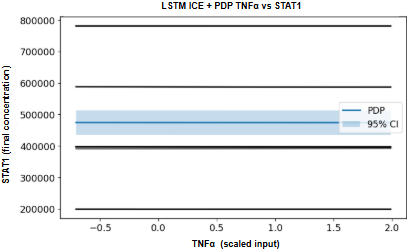} &
\includegraphics[width=0.31\linewidth]{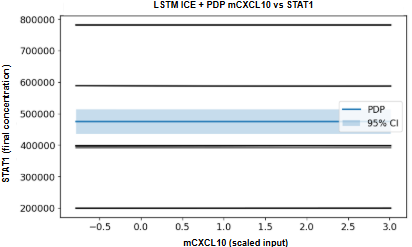} \\
\includegraphics[width=0.31\linewidth]{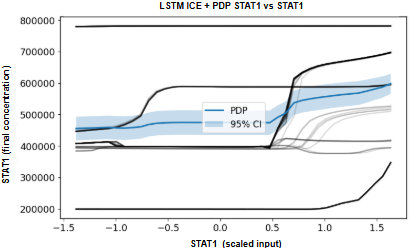} &
\includegraphics[width=0.31\linewidth]{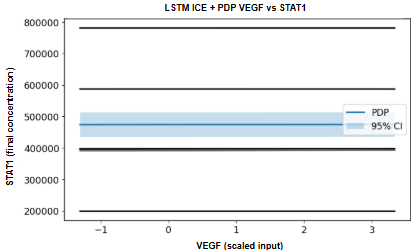} &
\includegraphics[width=0.31\linewidth]{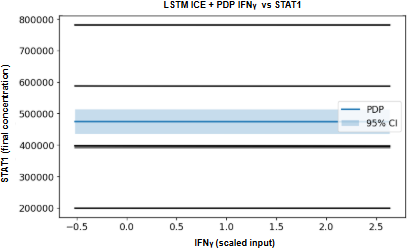}
\end{tabular}
\caption{ICE and PDP analysis for STAT1 prediction across key regulators. Gray lines represent individual conditional expectation (ICE) curves, the blue line shows the partial dependence plot (PDP), and the shaded region denotes the 95\% confidence interval. Strong nonlinear and threshold-like responses are observed for pro-inflammatory cytokines (iNOS, TNFa, mCXCL10), while VEGF and IFNG exhibit relatively weak marginal effects. Substantial variability across ICE curves highlights context-dependent regulatory dynamics.}
\label{fig}
\end{figure}

Across multiple features, the PDPs reveal distinct nonlinear and threshold-like behaviors. For strongly influential inflammatory features such as iNOS, TNF$\alpha$, and mCXCL10, the predicted STAT1 response remains relatively stable across a broad input range before increasing sharply beyond an intermediate feature value. These patterns suggest that the surrogate has learned switch-like activation dynamics in which STAT1 remains comparatively stable until sufficient inflammatory signaling accumulates.

In contrast, VEGF and IFNG display relatively flat population-averaged PDP responses. Their limited marginal effects do not indicate that these features are biologically unimportant. Rather, the substantial spread among their ICE curves demonstrates that their effects vary across individual simulated conditions and depend strongly on the surrounding regulatory state.

Several ICE trajectories also exhibit abrupt transitions around intermediate input values. These transitions indicate potential regime shifts in the learned dynamics, where small changes in a molecular feature can produce large changes in STAT1 output under particular cytokine environments. The widening confidence intervals near these transition regions further indicate increased heterogeneity and sensitivity. Together, the SHAP and ICE/PDP results distinguish globally influential features from context-dependent effects and identify input ranges in which the system may be especially responsive to perturbation.

\section{Discussion and Conclusions}

The results of this study demonstrate the effectiveness of surrogate neural models in capturing the complex, nonlinear behavior of macrophage polarization dynamics under varying microenvironmental conditions. The LSTM surrogate achieved strong prediction accuracy across thirteen key molecular markers, as evidenced by low training and validation losses and uniformly high $R^2$ scores. This confirms the model's capacity to generalize across the wide range of conditions generated through COPASI simulations, including varied hypoxia and regulatory input levels.

Polarization is governed by a complex network of signaling pathways, transcriptional programs, and metabolic rewiring. Canonical markers of M1 macrophages include iNOS, TNF$\alpha$, IL-12, and IFN-$\gamma$, while M2 macrophages express ARG1, IL-10, VEGF, and TGF$\beta$ \cite{lawrence2011transcr}. Environmental context, particularly oxygen availability, significantly shapes these phenotypic trajectories. Hypoxic conditions—common in tumor microenvironments, sites of chronic inflammation, and ischemic tissues—affect macrophage function through hypoxia-inducible factors such as HIF-1$\alpha$ and HIF-2$\alpha$, which modulate both metabolic and immune responses \cite{nizet2009hypox, palazon2014immun}.

\begin{figure*}[!htbp]
    \centering
    \includegraphics[
        width=\textwidth,
        height=0.69\textheight,
        keepaspectratio
    ]{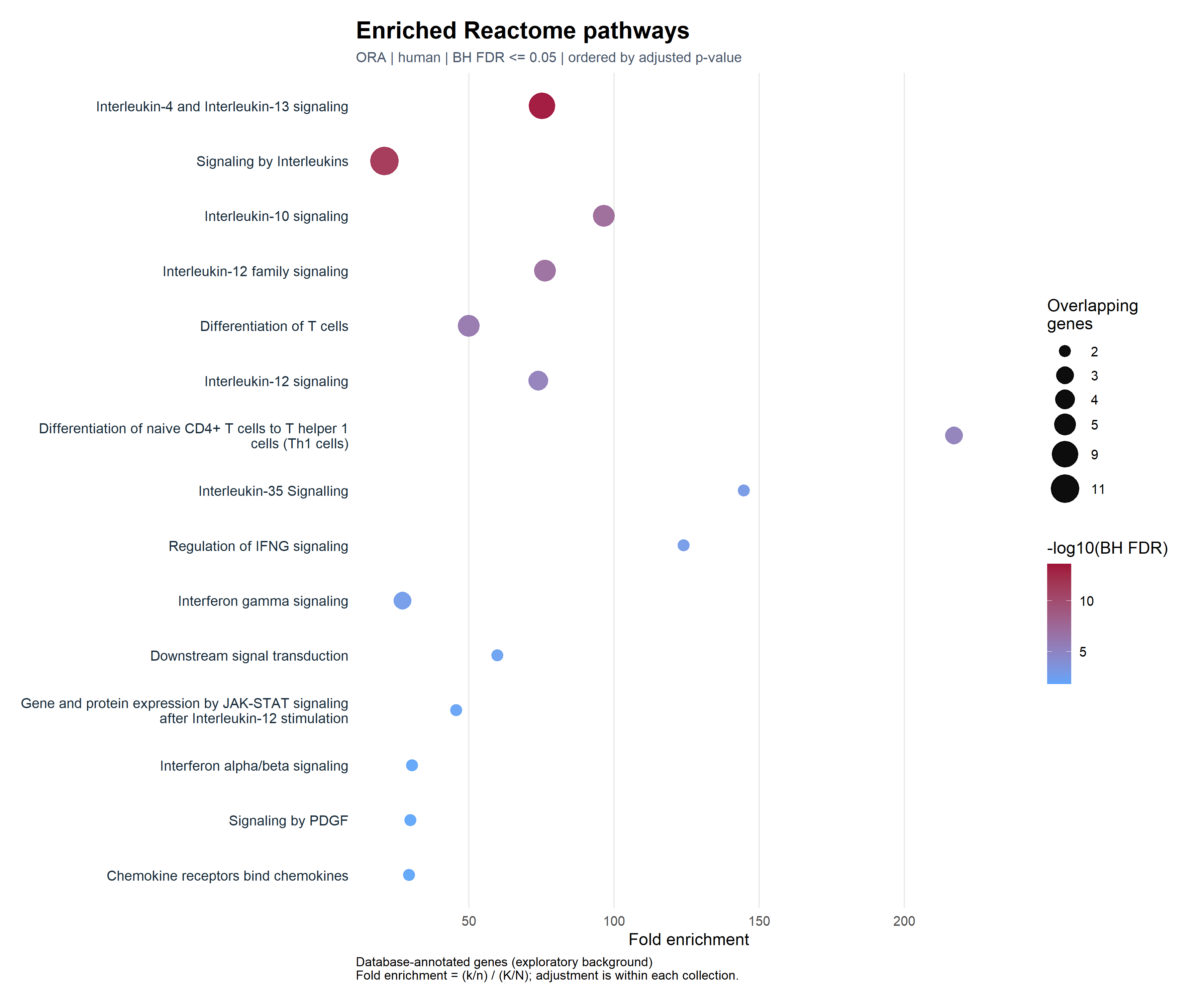}
    \caption{Enriched Reactome pathways associated with the top influential genes list. Displayed terms meet BH FDR $\leq 0.05$ and are ordered by adjusted $p$-value. Dot size indicates the number of overlapping input genes, and color represents $-\log_{10}(\mathrm{BH\ FDR})$. Fold enrichment is $(k/n)/(K/N)$, where $k$ and $n$ are the numbers of input genes in the term and tested input genes, respectively, and $K$ and $N$ are the corresponding background counts. The exploratory background comprises database-annotated human genes.}
    \label{fig:reactome_enrichment}
\end{figure*}

\begin{figure*}[!htbp]
    \centering
    \includegraphics[
        width=\textwidth,
        height=0.69\textheight,
        keepaspectratio
    ]{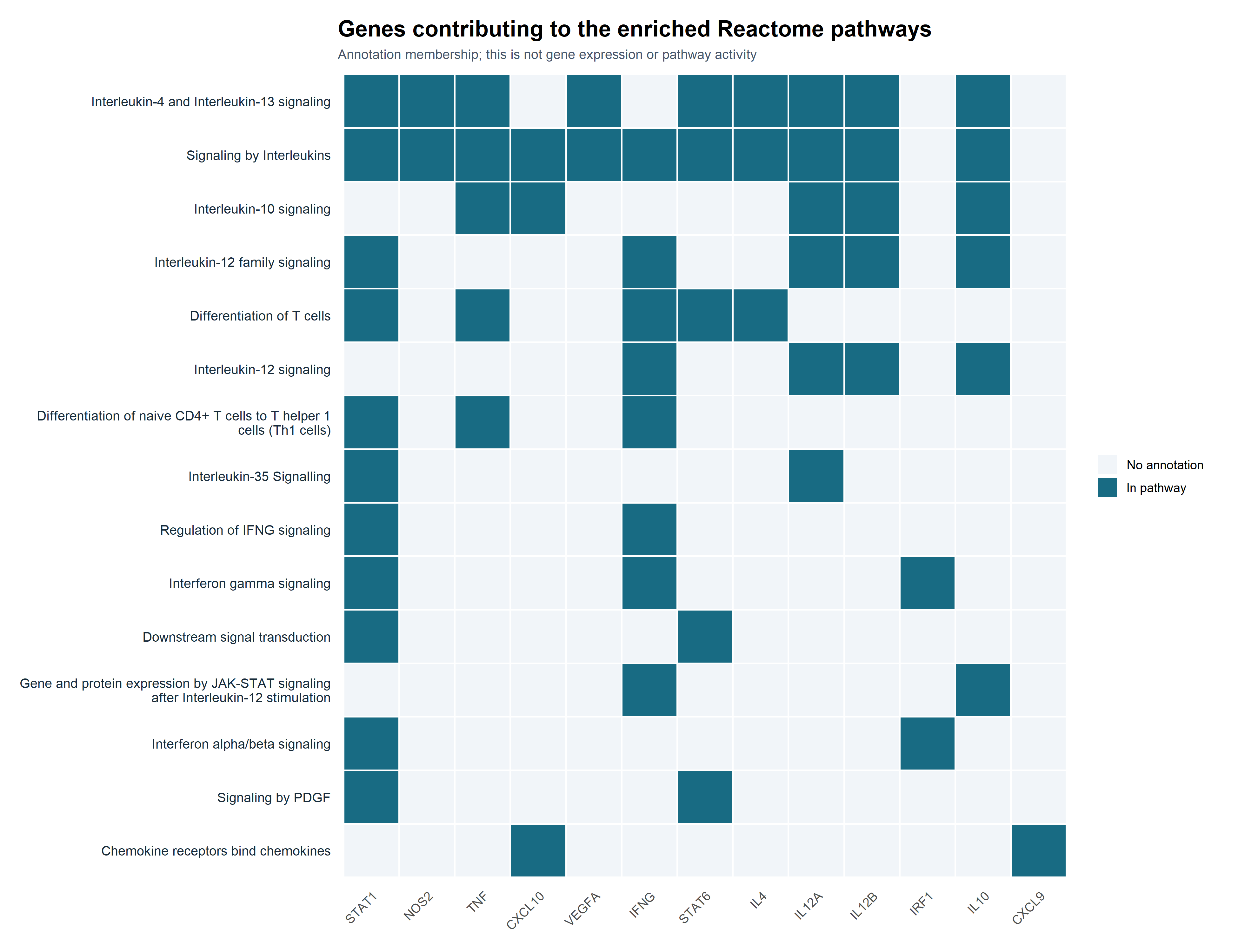}
    \caption{Input-gene membership in the enriched Reactome pathways. Filled cells indicate an annotation linking an input gene to a pathway; pale cells indicate no corresponding annotation. Gene symbols correspond to model variables including NOS2 (iNOS), TNF (TNF$\alpha$), CXCL10 (mCXCL10), VEGFA (VEGF), and IFNG (IFN$\gamma$). IL12 is represented by IL12A and IL12B; overlap counts therefore refer to mapped genes rather than the number of model features.}
    \label{fig:reactome_gene_pathway}
\end{figure*}

\begin{figure*}[!htbp]
    \centering
    \includegraphics[
        width=\textwidth,
        height=0.69\textheight,
        keepaspectratio
    ]{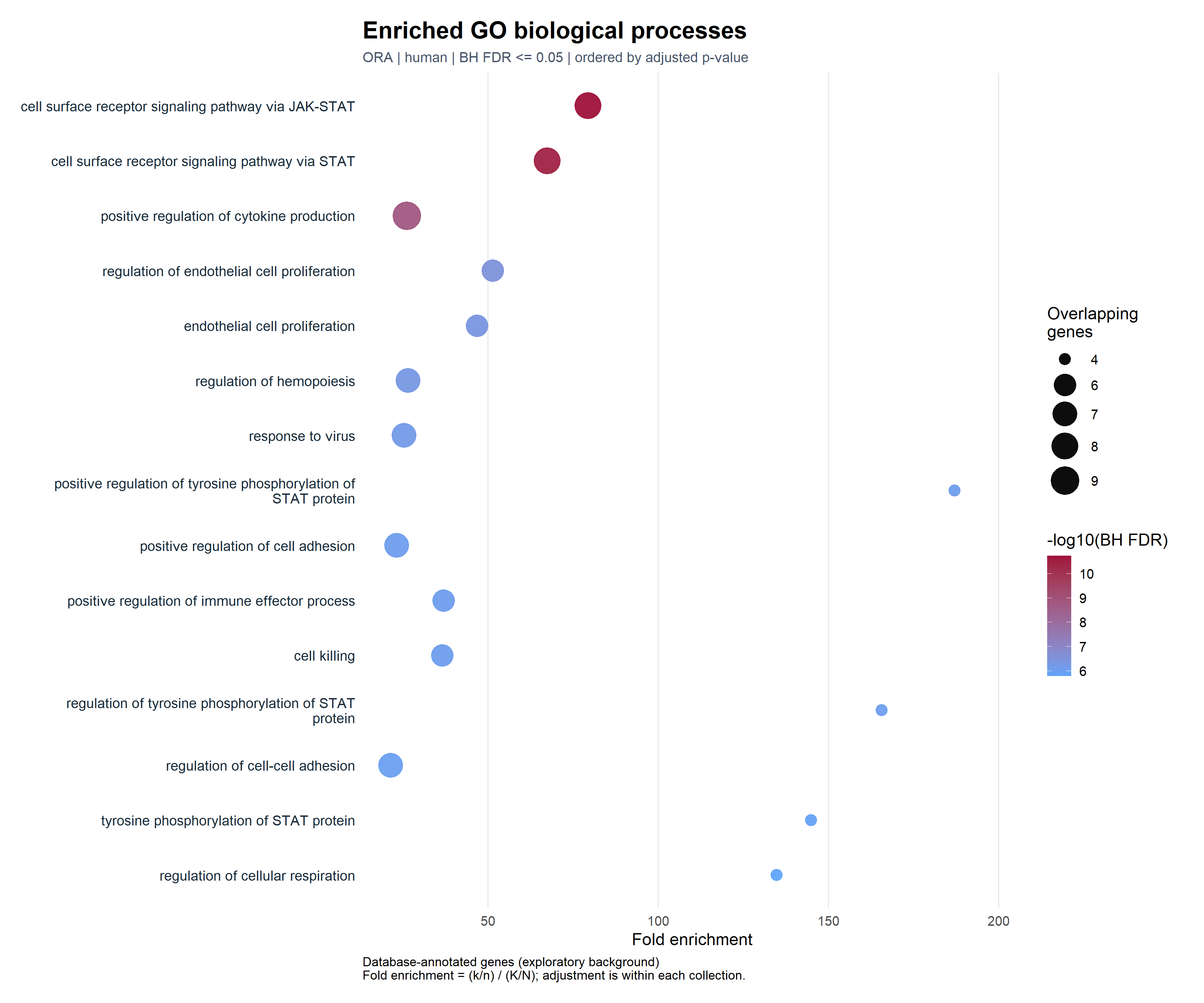}
    \caption{Enriched Gene Ontology biological processes associated with the analyzed gene list. Displayed terms meet BH FDR $\leq 0.05$ and are ordered by adjusted $p$-value. Dot size denotes the number of overlapping input genes, color denotes $-\log_{10}(\mathrm{BH\ FDR})$, and the horizontal axis shows fold enrichment relative to database-annotated human genes. Multiple-testing adjustment is performed within the GO collection. Enrichment indicates over-representation of annotations without establishing the direction of a biological response.}
    \label{fig:go_bp_enrichment}
\end{figure*}

In this study, a data-driven model was developed to learn macrophage polarization behavior from synthetic time-series data generated through simulations of the underlying molecular signaling network. SHAP quantified the contributions of individual molecular features to the surrogate predictions, while ICE and PDP analyses characterized the direction, nonlinearity, and context dependence of the learned relationships. To place the influential features in a broader biological context, the corresponding gene list was examined using Reactome pathway and Gene Ontology biological process enrichment analyses. The reported analyses use over-representation analysis (ORA), with Benjamini--Hochberg false discovery rate (BH FDR) correction within each annotation collection and a significance threshold of $0.05$. Reactome pathway enrichment, gene--pathway membership, and GO biological process enrichment are presented in Figures~\ref{fig:reactome_enrichment}--\ref{fig:go_bp_enrichment}.

Among the displayed Reactome results, \textit{Interleukin-4 and Interleukin-13 signaling} has the strongest statistical enrichment, followed by \textit{Signaling by Interleukins}, \textit{Interleukin-10 signaling}, and \textit{Interleukin-12 family signaling} (Figure~\ref{fig:reactome_enrichment}). These terms contain nine, eleven, five, and five overlapping input genes, respectively. Additional enriched terms include \textit{Interleukin-12 signaling}, \textit{Regulation of IFNG signaling}, \textit{Interferon gamma signaling}, \textit{Interferon alpha/beta signaling}, and \textit{Chemokine receptors bind chemokines}. Thus, the analyzed features span inflammatory, alternative-activation, and immunoregulatory cytokine networks. This breadth is consistent with the interacting signaling programs underlying macrophage plasticity \cite{sica2012, wang2014}.

The gene--pathway matrix identifies the annotations underlying these results (Figure~\ref{fig:reactome_gene_pathway}). STAT1 is shared across several interleukin and interferon terms, while IL4 and STAT6 contribute to the IL-4/IL-13 term. CXCL9 and CXCL10 account for the displayed chemokine-binding enrichment. The matrix also shows that inflammatory-associated genes, including NOS2 and TNF, occur within the IL-4/IL-13 annotation. Membership therefore identifies participation in an annotated pathway without specifying an activating or inhibitory role. Repeated genes across terms also mean that the enriched pathways are overlapping biological descriptions rather than independent signals, and their enrichment ranking should be interpreted separately from the SHAP feature ranking.

GO enrichment emphasizes \textit{cell surface receptor signaling pathway via JAK-STAT}, \textit{cell surface receptor signaling pathway via STAT}, and \textit{positive regulation of cytokine production} as the three most statistically enriched displayed processes (Figure~\ref{fig:go_bp_enrichment}). Terms involving STAT tyrosine phosphorylation further connect the gene list to cytokine-dependent signal transduction. Other enriched processes include response to virus, positive regulation of immune effector process, cell killing, regulation of cell adhesion, regulation of hemopoiesis, endothelial cell proliferation, and regulation of cellular respiration. These annotations place the SHAP findings within a broader network of immune signaling, vascular regulation, and metabolism.

The prominence of JAK--STAT and interferon-related annotations is consistent with STAT1 being the strongest SHAP predictor and with the generally positive contributions of high iNOS, TNF$\alpha$, and mCXCL10 values to predicted STAT1 output. These associations describe a coordinated inflammatory state in the simulated trajectories. Mechanistic modeling of LPS-stimulated macrophages has identified IFN-$\gamma$-dependent IRF1 activation as a major determinant of iNOS induction, with TNF$\alpha$ contributing to response dynamics \cite{salim2016}. Experimental evidence likewise establishes the importance of IRF1 for this output: stimulated IRF1-deficient murine macrophages produce little or no nitric oxide and exhibit markedly reduced iNOS mRNA \cite{kamijo1994}.

The TNF$\alpha$ and chemokine associations also have relevant biological precedents. TNF$\alpha$ can initiate an IRF1-dependent, IFN-$\beta$-mediated autocrine loop that sustains inflammatory gene expression and promotes delayed STAT1 expression in macrophages \cite{yarilina2008}. The enrichment of both type I and type II interferon annotations is also compatible with evidence that IFN-$\gamma$, as well as type I interferons, can induce promoter binding by the STAT1/STAT2/IRF9 complex in murine macrophages \cite{platanitis2019}. These studies provide context for the STAT1-associated inflammatory and chemokine features without establishing that every cited molecular mechanism is resolved by the surrogate.

The strong enrichment of IL-4/IL-13 and IL-10 signaling broadens this interpretation beyond inflammatory activation. In particular, the bidirectional SHAP distribution of STAT6 is compatible with context-dependent relationships between the IFN-$\gamma$/STAT1 and IL-4/STAT6 branches. The accompanying IL-12-related annotations place these relationships within a wider cytokine network. This interpretation accords with the balance among inflammatory, alternative-activation, and immunoregulatory programs described in macrophage polarization studies \cite{sica2012, wang2014}. Enrichment of these terms does not by itself identify which program is dominant in an individual simulated trajectory; that interpretation depends on the feature values and the SHAP and ICE/PDP responses.

The GO terms involving endothelial cell proliferation provide a functional context for the contribution of VEGF. High VEGF values have predominantly negative SHAP contributions, associating them with lower predicted STAT1 output under the sampled conditions. A relevant biological precedent is the delayed inhibition of VEGF-A translation by the IFN-$\gamma$-activated inhibitor of translation (GAIT) complex in human monocytic cells \cite{ray2007}. This mechanism supports a possible connection between interferon signaling and reduced angiogenic output, but it does not establish VEGF as an upstream inhibitor of STAT1. Similarly, the enrichment of regulation of cellular respiration provides a broader metabolic context for IRF1-associated effects. Silencing IRF1 reduces IRG1 expression and itaconic acid production in macrophages \cite{tallam2016}.

Because the analyzed features originate from a mechanistically selected macrophage network, enrichment against a broad database-annotated background should be interpreted as an exploratory assessment of biological coherence. The figures support a pathway-centered interpretation of the learned associations, while the direction and context dependence of the predictions remain grounded in SHAP and ICE/PDP rather than in enrichment significance alone.

The ICE and PDP results extend this interpretation by showing that feature importance does not imply a uniform effect across all simulated states. For iNOS, TNF$\alpha$, and mCXCL10, threshold-like response curves identify ranges in which inflammatory signaling may produce disproportionately large increases in STAT1 activity. By contrast, the flatter population-averaged effects of VEGF and IFNG conceal substantial differences among individual ICE trajectories. These findings support the conclusion that macrophage polarization is governed by a balance of signals whose effects depend on their relative abundance and cellular context rather than on the exclusive presence of individual markers.

The combination of feature ranking and response-curve analysis also provides a basis for future therapeutic research. The feature ranking and pathway analysis results should not be interpreted as a direct prescription for intervention, because SHAP values describe predictive contributions within the modeled system rather than establishing causal therapeutic effects. However, the rankings can prioritize candidate pathways for targeted perturbation studies. Highly ranked upstream regulators such as IFN$\gamma$, JAK signaling, TNF$\alpha$, and IL-4 may motivate experiments that test broad shifts between inflammatory and repair-associated states. Downstream regulators and readouts such as IRF1, iNOS, CXCL9, and mCXCL10 may support more selective interventions or provide biomarkers for determining whether a proposed treatment successfully alters pathway activity.

The nonlinear regions identified by ICE and PDP analysis may be especially useful for designing these experiments. Rather than testing only the presence or absence of a perturbation, future studies could concentrate measurements around predicted transition ranges where small changes in regulator abundance produce large changes in STAT1 output. Combinatorial perturbations could further determine whether these transitions are preserved, shifted, or eliminated when competing pathways are activated simultaneously. Experimental validation through cytokine administration, receptor blockade, gene knockdown, or pharmacological inhibition would help distinguish causal regulators from correlated markers of the same polarization state.

More broadly, the modeling paradigm is adaptable to different levels of biological resolution. It can be expanded by incorporating additional pathways, cell types, environmental variables, or intercellular signaling mechanisms to study novel regulatory networks and emergent immune behavior.

Conversely, the same framework can be narrowed to investigate a smaller pathway or therapeutic question with greater precision. A focused mechanistic model could densely sample a limited set of parameters, incorporate detailed receptor and transcription-factor kinetics, and generate fine-tuned explanations for a specific drug target or regulatory interaction. This flexibility permits the framework to serve either as a broad hypothesis-generation platform or as a targeted computational experiment for resolving subtle pathway-specific effects.

Overall, this study supports the utility of machine-learned surrogates not only for fast, high-fidelity prediction but also for biologically grounded hypothesis generation through explainable artificial intelligence. The ability to combine global feature attribution with individualized and population-averaged response curves provides a principled framework for uncovering regulatory hierarchies, nonlinear transition regions, and context-dependent effects in complex immune networks. Future work may extend this framework to stochastic or single-cell simulations, incorporate symbolic regression to extract compact analytical representations, and integrate experimental perturbation assays. These directions may help bridge computational predictions with wet-lab immunology and enable increasingly targeted modulation of macrophage responses in disease.

\end{document}